\documentclass[journal]{IEEEtran}
\usepackage{amsmath,amssymb,bm,booktabs,graphicx,array}
\usepackage{microtype}
\usepackage{xcolor}
\usepackage{url}
\usepackage{placeins}
\usepackage{tikz}
\usetikzlibrary{arrows.meta,positioning,calc}
\usepackage[hidelinks]{hyperref}
\hypersetup{pdftitle={Risk-Aware Safe Hierarchical Learning for Robust eMBB-URLLC Resource Allocation in O-RAN},
 pdfauthor={Mohammad Hussain, Dilara Gurer, Gokhan Kalem, Sinem Coleri}}
\newcommand{\U}{\mathrm U}
\newcommand{\E}{\mathrm E}
\newcommand{\cvar}{\operatorname{CVaR}}

\let\epsilon\varepsilon
\newcolumntype{L}[1]{>{\raggedright\arraybackslash}p{#1}}
\newcolumntype{C}[1]{>{\centering\arraybackslash}p{#1}}

\makeatletter
\g@addto@macro\normalsize{\setlength\jot{1.5pt}\setlength\abovedisplayskip{2.7pt plus 1pt minus 1pt}\setlength\belowdisplayskip{2.7pt plus 1pt minus 1pt}\setlength\abovedisplayshortskip{1pt plus 1pt}\setlength\belowdisplayshortskip{2pt plus 1pt minus 1pt}}
\makeatother
\begin{document}
\bstctlcite{IEEEexample:BSTcontrol}
\title{Risk-Aware Safe Hierarchical Learning for Robust eMBB--URLLC Resource Allocation in O-RAN}
\author{Mohammad~Hussain,
        Dilara~Gurer, Gokhan~Kalem,
        and~Sinem~Coleri,~\IEEEmembership{Fellow,~IEEE}%
\thanks{M. Hussain and S. Coleri are with the Department of Electrical and Electronics Engineering, Ko\c{c} University, Istanbul, T\"urkiye (email: \{mohussain25, scoleri\}@ku.edu.tr). D. Gurer and G. Kalem are with Turkcell (email: \{dilara.gurer, gokhan.kalem\}@turkcell.com.tr). This work is supported by the Scientific and Technological Research Council of T\"urkiye (T\"UB\.{I}TAK) 1711 Project AI-P5G \#3247019.}%
}

\maketitle

\begin{abstract}

Joint resource allocation for enhanced mobile broadband (eMBB) and ultra-reliable low-latency communication (URLLC) must satisfy stringent URLLC latency and reliability requirements while maintaining eMBB throughput under uncertain traffic and channel conditions. Existing learning-based allocators often optimize average aggregate performance without explicitly accounting for prediction errors, potentially resulting in severe service degradation for individual users. We propose a risk-aware safe hierarchical learning framework for Open Radio Access Networks (O-RAN) that learns resource-feasible allocations using service-cost penalties and a per-user URLLC service test. A non-real-time policy determines slice-level resource-block and power budgets, near-real-time policies allocate resource blocks and transmit power among users, and an execution-time safety layer checks per-user URLLC service and corrects allocations through bounded resource transfers when feasible. A deterministic per-user service test, derived using protected rates accounting for channel state information prediction errors and conformal upper bounds on traffic demand, ensures satisfaction of the probabilistic URLLC requirement. Service-specific conditional value-at-risk objectives prioritize users with the highest URLLC deadline-miss ratios and lowest eMBB rates. Experimental results show that, under nominal load, the proposed framework is the only evaluated method that satisfies both the URLLC deadline-miss target and worst-user eMBB rate criterion on average. It reduces the mean miss ratio by 64\% relative to the best baseline in this metric and achieves the highest worst-user eMBB rate.

\end{abstract}

\section{Introduction}\label{sec:introduction}

Fifth-generation (5G) and beyond networks must support heterogeneous services with distinct quality-of-service requirements over a shared radio access network (RAN). Enhanced mobile broadband (eMBB) requires high throughput, while ultra-reliable low-latency communication (URLLC) demands reliable packet delivery within stringent deadlines~\cite{3gpp22261reliability}. Allocating more resources to URLLC improves its ability to accommodate traffic bursts but reduces eMBB throughput, while smaller reservations increase resource utilization at the risk of missing URLLC deadlines~\cite{alsenwi2019}. Resource allocation must therefore balance reliable URLLC delivery, minimum per-user eMBB rates, and aggregate eMBB throughput.

The Open Radio Access Network (O-RAN) architecture supports adaptive resource allocation across multiple timescales~\cite{polese2023oran}. The non-real-time (non-RT) RAN Intelligent Controller (RIC) determines slice-level physical resource block (PRB) and power budgets, the near-real-time (near-RT) RIC assigns PRBs and transmit powers to users, and the distributed unit executes the allocations in radio slots. However, budgets and assignments remain fixed between updates while traffic demands and channel conditions evolve. Consequently, underestimated traffic demand or optimistic channel state information (CSI) predictions may cause service violations, potentially requiring resources to be reallocated between slices.

Deep reinforcement learning (DRL) has emerged as a promising approach to adaptive slicing and scheduling~\cite{setayesh2022,sohaib2024,safeslice2025}.
However, optimizing expected aggregate rewards can mask severe service degradation for individual users or time slots. Moreover, executing allocations without accounting for prediction uncertainty can expose URLLC users to repeated deadline misses and eMBB users to persistently low rates. These limitations motivate risk-aware resource allocation that explicitly accounts for service-specific tail performance and prediction uncertainty.

\subsection{Related Work}

\textit{Hierarchical resource allocation:} Hierarchical learning separates slice-level configuration from user-level scheduling. Setayesh et~al.~\cite{setayesh2022} use a
reinforcement-learning agent to determine slice bandwidth, power, and
numerology, followed by attention-based user schedulers, with a budget
reward based on throughput and service satisfaction. Akyildiz
et~al.~\cite{akyildiz2024} coordinate inter-slice and intra-slice agents
at each transmission time interval using quantized queues and
channel-based user groups. Qiao et~al.~\cite{qiao2025} use separate agents for slice numerology, radio and computing budgets over long slots and user allocation over short slots. Bao et~al.~\cite{bao2026agentic} adapt PRB budgets using URLLC violation, fallback statistics and eMBB backlog, combining learned scheduling with delay-triggered fallback.
Although these approaches demonstrate the benefits of hierarchical control, they neither explicitly optimize service-specific tail performance nor verify allocations against probabilistic URLLC service requirements under joint traffic-demand and CSI uncertainty.

\textit{Risk-aware scheduling:} Expected throughput or reward can obscure severe service degradation. Shi et~al.~\cite{shi2024puncturing} minimize the sum of eMBB throughput-loss ratios, with a learned puncturing policy that maximizes expected eMBB rate under URLLC delay and reliability constraints. Sohaib et~al.~\cite{sohaib2024} maximize expected discounted reward combining aggregate eMBB rate with an adaptive URLLC reliability penalty. Neither formulation includes an explicit tail-risk measure. Alsenwi et~al.~\cite{alsenwi2019} combine proportional-fair expected eMBB rates with a conditional value-at-risk (CVaR) term for aggregate eMBB rate under puncturing. Zhang et~al.~\cite{zhang2024} maximize CVaR of discounted long-term reward, combining URLLC packet nondrop probability and aggregate eMBB throughput, subject to a CVaR constraint on long-term cost derived from the URLLC outage requirement. These risk-sensitive formulations focus on aggregate eMBB rate or discounted returns, rather than separately modeling eMBB rate shortfalls across time and users, URLLC delay tails, and per-user deadline-miss ratios.

\textit{Prediction uncertainty:} Traffic and channel prediction methods have been developed to improve resource allocation under uncertainty.  Kavehmadavani et~al.~\cite{kavehmadavani2023} use long short-term memory (LSTM) models for traffic prediction in O-RAN, but point predictions do not quantify possible demand underestimation. Cohen et~al.~\cite{cohen2023} employ online conformal demand prediction to adapt uplink URLLC allocations to a long-run frame-level reliability target. 
For channel uncertainty, Hussain et~al.~\cite{hussain2026csi} reduce the computational cost of CSI prediction. However, CSI prediction errors can still lead to optimistic rate estimates. Mehrnia and Coleri~\cite{mehrnia2022confidence}
incorporate confidence intervals for channel-tail parameters into extreme-value-based rate selection.
Sariisik and Coleri~\cite{sariisik2026robust} use Bernstein's inequality to bound channel power under CSI uncertainty and optimize blocklength through knapsack-based DRL with model-based action correction. However, their framework assumes known traffic demand from fixed-size periodic updates. Gharehgoli
et~al.~\cite{gharehgoli2023} jointly model traffic demand and CSI uncertainty using recurrent reinforcement learning for end-to-end slicing. However, these studies do not jointly incorporate calibrated traffic-demand bounds and probabilistic per-user CSI prediction-error bounds to verify URLLC service feasibility during fixed allocation intervals.

\textit{Safe learning:} Safe learning methods incorporate constraints during policy optimization or correct actions before execution. Constrained policy optimization controls expected costs during learning. Zangooei et~al.~\cite{cmars2024} constrain expected slice-level service violations in multi-agent slicing. In contrast, safety layer corrects individual actions before execution. Dalal et~al.~\cite{dalal2018} project policy outputs onto learned linear constraints through quadratic programming. SafeSlice~\cite{safeslice2025} combines a risk-sensitive latency reward with a learned safety model that corrects slice bandwidth shares for the predicted latency compliance. Mahalakshmi and Markkandan~\cite{mahalakshmi2026tail} combine a URLLC tail-delay penalty in proximal policy optimization (PPO) with a safety layer that transfers mMTC and V2X resources to URLLC when satisfaction falls below a threshold. In O-RAN, MAREA~\cite{adamuz2025marea} uses a martingale-based delay model to determine URLLC reservations, with real-time adjustments triggered by queue anomalies. Ali et~al.~\cite{ali2024} use model-derived optimality conditions to reduce the number of constraint violations in ultra-reliable wireless control. Gunes Reyhan and Coleri~\cite{reyhan2025safe} use a teacher--student safety mechanism that replaces infeasible actions with the nearest feasible actions to satisfy peak age-of-information violation, power, and schedulability constraints during learning. Resource preemption provides a related mechanism for adjusting allocations at execution time, although, unlike constraint-aware safety layers, it does not necessarily verify satisfaction of explicit service requirements. Anand et~al.~\cite{anand2020} study URLLC puncturing of eMBB resources under different rate-loss models. Saggese et~al.~\cite{saggese2021} use DRL to select the punctured eMBB resources, with limited puncturing per codeword. However, these works do not incorporate explicit traffic-demand and CSI uncertainty bounds into per-user service verification nor determine the minimum required PRB transfer.

Table~\ref{tab:related} compares the closest related studies across five dimensions: two-timescale budgets, per-service tail risk, demand prediction-error bound, CSI prediction-error bound, and safety layer. Existing methods address at most three of these dimensions, some only partially, whereas the proposed framework integrates all five.

\begin{table}[t]
\centering
\caption{Comparison with related studies (\checkmark: addressed;
$\circ$: partially addressed; --: not addressed).}
\label{tab:related}
\footnotesize\renewcommand{\arraystretch}{1.0}
\setlength{\tabcolsep}{0.8pt}
\setlength{\heavyrulewidth}{0.6pt}
\setlength{\lightrulewidth}{0.3pt}
\begin{tabular*}{\columnwidth}{@{\extracolsep{\fill}}lccccc@{}}
\toprule
\textbf{Ref.} & \shortstack{\textbf{Two-}\\\textbf{timescale}\\\textbf{budgets}} & \shortstack{\textbf{Per-}\\\textbf{service}\\\textbf{tail risk}} & \shortstack{\textbf{Demand}\\\textbf{pred.-error}\\\textbf{bound}} & \shortstack{\textbf{CSI}\\\textbf{pred.-error}\\\textbf{bound}} & \shortstack{\textbf{Safety}\\\textbf{layer}}\\
\midrule
\cite{setayesh2022,qiao2025} & \checkmark & -- & -- & -- & -- \\
\cite{bao2026agentic} & \checkmark & -- & -- & -- & \checkmark \\
\cite{adamuz2025marea} & \checkmark & -- & $\circ$ & -- & $\circ$ \\
\cite{alsenwi2019,zhang2024} & -- & $\circ$ & -- & -- & -- \\
\cite{safeslice2025,mahalakshmi2026tail} & -- & $\circ$ & -- & -- & \checkmark \\
\cite{cohen2023} & -- & -- & \checkmark & -- & -- \\
\cite{gharehgoli2023} & -- & -- & $\circ$ & $\circ$ & -- \\
\cite{mehrnia2022confidence} & -- & -- & -- & $\circ$ & -- \\
\cite{sariisik2026robust} & -- & -- & -- & \checkmark & \checkmark \\
\midrule
\textbf{Ours} & \checkmark & \checkmark & \checkmark & \checkmark & \checkmark \\
\bottomrule
\end{tabular*}
\end{table}

\subsection{Main Contributions}

We propose a risk-aware, safe hierarchical learning framework for joint eMBB--URLLC resource allocation in O-RAN under traffic-demand and CSI uncertainty. The framework combines service-specific risk objectives, per-user uncertainty bounds for both traffic demand and CSI, and an execution-time safety layer that verifies and, when necessary, corrects resource allocations. The main contributions are as follows:
\begin{itemize}

\item We formulate a two-timescale resource-allocation problem that jointly optimizes slice-level PRB and power budgets at the non-RT RIC and per-PRB user assignments and transmit powers at the near-RT RIC. The formulation incorporates aggregate eMBB throughput, minimum per-user eMBB rates, long-run URLLC deadline reliability, resource constraints, and per-user probabilistic URLLC service requirements.

\item We introduce CVaR-based service-specific risk-aware objectives to model performance degradation overlooked by average rewards. Separate eMBB risk terms penalize extreme rate deficits across time and users, while URLLC risk terms emphasize users with the highest deadline-miss ratios.

\item We derive a deterministic per-user service test that jointly accounts for traffic-demand and CSI prediction errors. Under the assumed CSI prediction-error model and demand-coverage condition, satisfying this test at each assignment boundary is sufficient to meet the marginal per-user probabilistic URLLC service requirement. 

\item We develop an execution-time safety layer that verifies resource allocations before execution and, when a feasible correction exists, transfers the minimum number of PRBs from eMBB to URLLC subject to predefined transfer limits. The CVaR objectives guide statistical tail performance during training, whereas the service test and safety layer act on individual allocations before execution.

\item Experimental results demonstrate that, at nominal load, the proposed framework is the only evaluated method that satisfies both the URLLC deadline-miss target and the worst-user eMBB rate criterion on average. It reduces the mean URLLC deadline-miss ratio by 64\% compared with the best-performing baselines in this metric and achieves the highest worst-user eMBB rate.

\end{itemize}

The remainder of this paper is organized as follows. Section~\ref{sec:system_model} presents the system model, Section~\ref{sec:problem} formulates the resource-allocation problem, Section~\ref{sec:protection} derives the deterministic service test, and Section~\ref{sec:hierarchy} presents the hierarchical framework and safety layer. Section~\ref{sec:eval} provides the performance evaluation, and Section~\ref{sec:conclusion} concludes the paper.

\section{System Model and Assumptions}\label{sec:system_model}

We consider the downlink of a single O-RAN cell serving URLLC
and eMBB users on a shared carrier. Let \(\mathcal{U}^{\U}\) and \(\mathcal{U}^{\E}\) denote the sets of URLLC and eMBB users, indexed by \(i\) and \(j\), respectively. Let \(u\) denote an arbitrary user in \(\mathcal{U}=\mathcal{U}^{\U}\cup\mathcal{U}^{\E}\). 
The carrier comprises $N$ PRBs
of equal bandwidth, indexed by $r\in\mathcal N=\{1,\ldots,N\}$. The total and per-PRB transmit powers are limited to \(P_{\max}^{\mathrm{cell}}\) and \(P_{\max}^{\mathrm{PRB}}\), respectively.

\subsection{O-RAN Control Layers and Timescales}\label{subsec:arch}
 
The O-RAN architecture separates the base station into an open radio unit (O-RU), an open distributed unit (O-DU) hosting the medium access control scheduler, and an open central unit (O-CU)~\cite{polese2023oran}. The
non-RT RIC resides within the service management and orchestration
(SMO) framework. It hosts rApps operating on non-real-time timescales and communicates policies to the near-RT RIC over the A1
interface. The near-RT RIC hosts xApps that control O-DU resource allocation over the E2 interface on near-real-time timescales. Measurements are reported through the E2 and O1 interfaces.

The proposed controller uses both the non-RT and the near-RT control loops, so it
operates on two decision timescales above the radio-slot timescale. Layer~1 is implemented as an rApp and determines the slice budgets. Layers 2A, 2B, and 3 are implemented as xApps and respectively allocate PRBs and transmit powers to URLLC users, allocate the remaining PRBs and their transmit powers to eMBB users, and correct the resulting allocation when necessary. The O-DU executes allocations in radio slots of duration $\tau$, indexed by $k$. The xApps update the user allocation once per
\emph{assignment interval} of $K$ slots. Interval $m$
starts at the \emph{assignment boundary} $k_m=mK$ and contains the slots
$\mathcal K_m=\{k_m,\ldots,k_m+K-1\}$. The rApp updates the slice
budgets once per \emph{orchestration period} of $M$ assignment intervals, indexed
by $t$. The assignment interval and orchestration period are chosen to match the respective control timescales. Each user allocation remains fixed for \(K\) slots, during which traffic arrivals and channel conditions continue to evolve.

\begin{figure*}[t]
\centering
\includegraphics[width=0.88\textwidth]{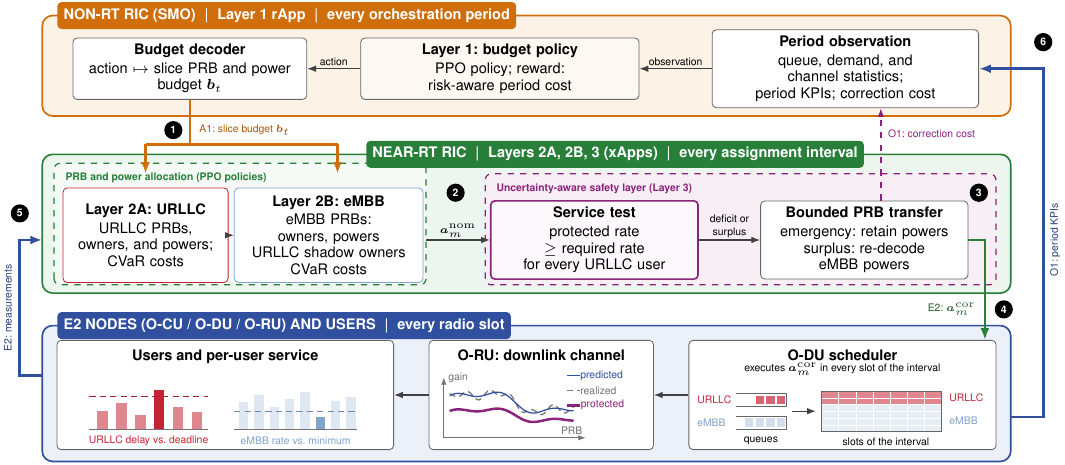}
\caption{Proposed risk-aware safe hierarchical framework in
O-RAN; circled numbers give the order of one control cycle.}
\label{fig:architecture}
\end{figure*}

Fig.~\ref{fig:architecture} illustrates the control cycle. At each
orchestration boundary, Layer~1 sends the slice budgets to the near-RT RIC over
the A1 interface. At each assignment boundary, Layer~2A
selects the URLLC PRBs and assigns them to URLLC users with their transmit powers, while Layer~2B
assigns the remaining PRBs to eMBB users with their transmit powers. Layer~3 then checks the per-user URLLC service test and applies bounded PRB transfers to address service deficits or release surplus resources. The corrected allocation is sent to the O-DU over E2 and executed in every slot of the interval. E2 measurements update the states of
Layers~2A, 2B, and~3, while period-level key performance indicators
(KPIs) and correction costs are reported to
Layer~1 over O1.

\subsection{Slice Budgets and User Allocation}\label{subsec:resources}

At the non-RT timescale, the budget $\bm b_t=(N_t^{\U},P_t^{\U})$ specifies the PRBs and power reserved for URLLC in period $t$. The remaining resources are assigned to the eMBB slice. Accordingly,
\begin{align}
 &N_t^{\U}+N_t^{\E}=N,\qquad P_t^{\U}+P_t^{\E}=P^{\rm cell}_{\max},
 \label{eq:budget_sum}\\
 &N_t^{\rho}\in[N_{\min}^{\rho},N_{\max}^{\rho}],\qquad
 P_t^{\rho}\in[P_{\min}^{\rho},P_{\max}^{\rho}],
 \label{eq:budget_set}
\end{align}
where $N_t^{\E}$ and $P_t^{\E}$ denote the PRB and power budgets of
eMBB, and, for each slice $\rho\in\{\U,\E\}$,
$[N_{\min}^{\rho},N_{\max}^{\rho}]$ and $[P_{\min}^{\rho},P_{\max}^{\rho}]$
are configured bounds that keep the slice from being starved.

At the near-RT timescale, the controller assigns each PRB to at most one user and specifies its transmit power. Let
$\bm a_k=(\bm x_k,\bm p_k)$ denote the allocation in slot $k$, where $x_{u,r,k}=1$ if PRB
$r$ is assigned to user $u$, and $p_{r,k}$ is its transmit power. The assignment and power variables satisfy
\begin{equation}
 \begin{aligned}
 &\textstyle\sum_{u\in\mathcal U}x_{u,r,k}\le1,\qquad x_{u,r,k}\in\{0,1\},\\
 &0\le p_{r,k}\le P^{\rm PRB}_{\max}\textstyle\sum_{u\in\mathcal U}x_{u,r,k},
 \end{aligned}
 \label{eq:ownership}
\end{equation}
for every PRB $r$ and slot $k$. Thus, an unassigned PRB transmits no power.

In interval $m$, Layers~2A and~2B
produce a nominal allocation $\bm a_m^{\rm nom}$, reserving a URLLC
subset $\mathcal N_m^{\U}\subseteq\mathcal N$ with $|\mathcal N_m^{\U}|=N_t^{\U}$, where \(t\) is the period containing interval \(m\).
Layer~3 converts this allocation into the corrected allocation $\bm a_m^{\rm cor}$, which is
executed throughout the interval:
\begin{equation}
 \bm a_k=\bm a_m^{\rm cor},\qquad k\in\mathcal K_m.
 \label{eq:hold}
\end{equation}

\subsection{Bounded Transfers Between Slices}\label{subsec:transfers}

Prediction errors can leave URLLC users with insufficient or excess resources. Layer~3 corrects the nominal allocation through bounded inter-slice PRB transfers.

In interval $m$, the safety layer may transfer $n_m^{\rm emg}$ PRBs from eMBB to URLLC
(\emph{emergency transfer}) or $n_m^{\rm rel}$ PRBs from URLLC to eMBB
(\emph{surplus release}), but not both. Let $\mathcal N_m^{\U,\rm cor}$ denote the corrected URLLC PRB set, and define $n_m^{\rm emg}=|\mathcal N_m^{\U,\rm cor}\setminus\mathcal N_m^{\U}|$ and $n_m^{\rm rel}=|\mathcal N_m^{\U}\setminus\mathcal N_m^{\U,\rm cor}|$. The owners and powers of all other PRBs remain fixed, except that surplus release recomputes the eMBB powers. These transfers are limited as
\begin{equation}
 \begin{aligned}
 0\le n_m^{\rm emg}&\le\min\{N_{\rm emg},\,N_t^{\E}-N_{\rm fl}^{\E}\},\\
 0\le n_m^{\rm rel}&\le\min\{N_{\rm rel},\,N_t^{\U}-N_{\rm fl}^{\U}\},\\
 n_m^{\rm emg}n_m^{\rm rel}&=0,
 \end{aligned}
 \label{eq:transfer}
\end{equation}
where $N_{\rm emg}$ and $N_{\rm rel}$ are the transfer caps and
$N_{\rm fl}^{\U}$ and $N_{\rm fl}^{\E}$ are the
minimum numbers of PRBs that each slice keeps after a transfer. Power may be redistributed along with the transferred PRBs without increasing the cell-wide power limit:
\begin{equation}
 p_k^{\U}\le P_t^{\U}+\Delta P_m,\qquad
 p_k^{\E}\le P_t^{\E}-\Delta P_m,\qquad k\in\mathcal K_m,
 \label{eq:power}
\end{equation}
where $\Delta P_m\in[-P_{\rm rel},P_{\rm emg}]$ is the change in the URLLC power grant. During an emergency transfer, $\Delta P_m=[p_{k_m}^{\U}-P_t^{\U}]^+$ is the corrected URLLC power in excess of its grant. Here, $P_{\rm emg}$ and $P_{\rm rel}$ are
the power-transfer caps, and $p_k^{\rho}$ is the total transmit power of
the PRBs of slice $\rho$ in slot $k$. Positive and negative $\Delta P_m$ increase the URLLC and eMBB grants, respectively. The adjusted power budgets must remain nonnegative.

\subsection{Channel Model and CSI Prediction}\label{subsec:csi}

Let $g_{u,r,k}$ denote the realized channel of user $u$ on PRB $r$ in slot $k$, which depends on propagation, mobility, shadowing, and fading. Since CSI reports reach the controller with a delay, the near-RT controller predicts this channel as $\widehat g_{u,r,k}$. The predicted channel is given by
\begin{equation}
 \widehat g_{u,r,k}=g_{u,r,k}+\Delta g_{u,r,k},
 \label{eq:channel}
\end{equation}
where $\Delta g_{u,r,k}$ denotes the CSI prediction error. Let $\mathcal F_k$ denote the information available to the controller when it selects the allocation of slot $k$, including the available CSI predictions, delayed observations, queue and traffic information, and the applicable slice budgets. The selected allocation is determined from the information in $\mathcal F_k$. Conditional on $\mathcal F_k$, the real and imaginary parts of $\Delta g_{u,r,k}$ are independent zero-mean Gaussian variables with standard deviation $c_{\rm CSI}|\widehat g_{u,r,k}|$, where $c_{\rm CSI}$ is the CSI prediction-error parameter. The controller observes $\widehat g_{u,r,k}$ but not the realized channel $g_{u,r,k}$, and uses the prediction-error statistics to back off the magnitude of the predicted CSI. The realized channel determines the achievable rate.

The signal-to-noise ratio (SNR) of user $u$ on PRB $r$ is given by
\begin{equation}
 \gamma_{u,r,k}=p_{r,k}|g_{u,r,k}|^2/P_{\rm noise},
 \label{eq:sinr}
\end{equation}
where $P_{\rm noise}$ is the noise power per PRB. We assume link adaptation at the O-DU based on the realized SNR, so a CSI prediction error does not directly cause a decoding error but can cause the rate anticipated from the predicted CSI to exceed the achievable rate. With $N_{\rm RE}$ data resource elements per PRB and slot, the achievable rate of user $u$ in slot $k$ is given by
\begin{equation}
 R_{u,k}=\frac{N_{\rm RE}}{\tau}\sum_{r\in\mathcal N}x_{u,r,k}\,s(\gamma_{u,r,k}),
 \label{eq:rates}
\end{equation}
where $s(\gamma)$ is the spectral efficiency selected by link adaptation. Specifically, it is the largest spectral efficiency in the channel quality indicator (CQI) table~\cite{3gpp38214} that does not exceed $\log_2(1+\gamma/\Gamma)$, or zero if no entry satisfies this condition, with the configured linear-scale capacity gap $\Gamma$.
\subsection{Traffic Model and Demand Prediction}\label{subsec:traffic}

We model the arrivals of each user as a slotted two-state
Markov-modulated Poisson process (MMPP), which
alternates between a calm and a burst regime. The regime switches between these states with fixed per-slot transition probabilities and is not observed by the controller. Given the regime, the
number of packets of a user in a slot is Poisson distributed, with a higher mean in
the burst regime. Both means are scaled by a slice-specific load factor
that cycles through predefined levels, each maintained for a fixed number of slots.
The packets of each user have a fixed size, and let $A_{u,k}$ denote the
number of bits arriving for user $u$ in slot $k$.

Because the allocation remains fixed throughout an assignment interval, the controller
must anticipate the traffic arriving during that interval. At boundary $k_m$, a demand predictor estimates the per-slot arrivals $\widehat A_{u,k_m}$ of every
user from the arrivals observed before $k_m$, and this prediction is used
for every slot of the interval. For URLLC users, a conformal demand margin is added to account for possible underprediction of demand.

\subsection{Queues and URLLC Packets}\label{subsec:queues}

Let $Q_{u,k}$ denote the backlog of user $u$ before the arrivals of slot
$k$, and let $Q_u^{\max}$ denote its buffer capacity. The backlog evolves as
\begin{equation}
 Q_{u,k+1}=\min\{[Q_{u,k}+A_{u,k}-\tau R_{u,k}]^+,Q_u^{\max}\},
 \label{eq:queue}
\end{equation}
where $[\cdot]^+=\max\{\cdot,0\}$. The buffer limit is applied to the residual backlog after service, so that bits served in a slot free buffer space for the arrivals of that slot, and any excess bits are dropped.

URLLC packets are tracked individually and served in first-in, first-out (FIFO) order. A
packet of user $i$ must be delivered within the deadline $D_i$, which
includes a fixed processing delay $\phi_i$. If it arrives in slot $k$
and completes in slot $k'$, its delay is given by
\begin{equation}
 L=\phi_i+(k'-k+1)\tau,
 \label{eq:fifo_delay}
\end{equation}
so it meets the deadline exactly when $k'\le k+h_i$ with
$h_i=\lfloor(D_i-\phi_i)/\tau\rfloor-1$, where $D_i-\phi_i$ is a positive integer multiple of $\tau$. A packet \emph{fails} if it has
not completed transmission by the end of slot $k+h_i$ or if any of its bits are dropped because of buffer overflow. A packet that misses its deadline is counted as failed but remains in the queue and is still served in FIFO order. Similarly, when overflow drops only part of a packet, its remaining bits stay in the FIFO queue. The queue evolution~\eqref{eq:queue} is therefore not affected by deadline expiry. 

For reliability accounting, a packet becomes
\emph{due} after slot $k+h_i$ has elapsed. Let $V_t$ denote the deadline-miss ratio of period $t$, i.e.,
the fraction of the URLLC packets falling due in period $t$ that fail,
with $V_t=0$ when no packet falls due. Similarly, let $V_{i,t}$ denote the deadline-miss ratio
of user $i$ in period $t$.

\subsection{Service Targets}\label{subsec:targets}

For eMBB user $j$, let $B_{j,k}$ denote the delivered rate in slot
$k$, i.e., the bits it receives in slot $k$ divided by $\tau$, and 
$B_{j,m}$ and $B_{j,t}$ denote its averages over assignment interval $m$ and orchestration period $t$, respectively.

The policy $\pi$ uses information available at each decision boundary. Let $F_t$ and $A_t^{\rm due}$ denote the failed packets among those falling due and all packets falling due in period $t$. Define the long-run packet-weighted ratio $\bar V(\pi)=\limsup_{T\to\infty}\sum_{t<T}\mathbb E_\pi[F_t]/\sum_{t<T}\mathbb E_\pi[A_t^{\rm due}]$~\cite{3gpp22261reliability}. A bar otherwise denotes the long-run period average under $\pi$, including traffic and channel realizations. The period miss ratio $V_t$ is used during training. The service
targets require
\begin{equation}
 \bar V(\pi)\le\epsilon_V,\qquad \bar B_j(\pi)\ge B_{\min},\quad j\in\mathcal U^{\E},
 \label{eq:service_targets}
\end{equation}
where $\epsilon_V$ is the target URLLC deadline-miss ratio and $B_{\min}$ is the minimum long-run eMBB rate.

\subsection{Probabilistic URLLC Service Requirement}\label{subsec:requirement}

In addition to the long-run service targets, the formulation imposes a per-user probabilistic URLLC service requirement at each assignment boundary.
At boundary $k_m$, URLLC user $i$ must serve $W_{i,m}=Q_{i,k_m}+A_{i,k_m}$
bits, which include the existing backlog and the arrivals of the boundary slot. We express this
need as a required rate for $W$ bits to serve, with three terms, each of
which protects against a different way of missing deadlines. First, a sustained rate $R$ should serve the $W$ bits within the $h_i+1$ slots available to a packet arriving in the boundary slot, which requires $R\ge W/((h_i+1)\tau)$. Second, the
backlog must not build up during a burst, so it must fall to a safe level
$Q_i^{\rm safe}$ within one slot, which requires
$(W-Q_i^{\rm safe})/\tau$. Third, packets close to their deadlines must
be served in time even when the total backlog is small, which requires
$\max_{\ell\ge1}W_{i,m}^{\rm due}(\ell)/(\ell\tau)$, where $W_{i,m}^{\rm due}(\ell)$ is the number of bits already queued at the boundary that are due within the next $\ell$ slots, with overdue bits counted at $\ell=1$. The required rate is
therefore given by
\begin{equation}
 \begin{aligned}
 R_{i,m}^{\rm req}(W)=\max\Big\{&
 \frac{W}{(h_i+1)\tau},\
 \frac{W-Q_i^{\rm safe}}{\tau},\\
 &\max_{\ell\ge1}\frac{W_{i,m}^{\rm due}(\ell)}{\ell\tau}\Big\},
 \end{aligned}
 \label{eq:requirement}
\end{equation}
which is nondecreasing in the bits to serve $W$. To account for uncertainty in boundary-slot arrivals and channel conditions, we impose the marginal chance constraint
\begin{equation}
 \Pr\bigl\{R_{i,k_m}\ge R_{i,m}^{\rm req}(W_{i,m})\bigr\}\ge1-\epsilon_i,
 \label{eq:chance_def}
\end{equation}
where $\epsilon_i\in(0,1)$ is the maximum allowable probability that
user $i$ falls below its required rate at a boundary. 
The controller does not know the arrivals and the channel of the boundary slot when it decides. It therefore uses the demand prediction and the CSI prediction, each protected by a margin against its prediction error.

\subsection{Cost Model}\label{subsec:cost_model}
The period cost balances URLLC delay and failures, eMBB throughput and rate shortfalls, and allocation corrections and budget changes. To capture severe service degradation that average performance can conceal, several components use CVaR.

At level $\beta\in[0,1)$, CVaR is the mean of the worst $(1-\beta)$ fraction
of the samples. For a nonempty finite multiset $\mathcal{Y}$, in which repeated values
are counted separately, it is given by the Rockafellar--Uryasev
representation~\cite{rockafellar_uryasev2000}
\begin{equation}
\operatorname{CVaR}_{\beta}(\mathcal{Y})
=
\min_{\zeta\in\mathbb{R}}
\left\{
\zeta
+
\frac{1}{(1-\beta)|\mathcal{Y}|}
\sum_{y\in\mathcal{Y}} [y-\zeta]^+
\right\},
\label{eq:cvar}
\end{equation}
where a minimizing $\zeta$ is the empirical $\beta$-quantile. Each CVaR
cost uses its own level $\beta$.

\paragraph{URLLC terms}
The URLLC terms are designed to \emph{minimize packet delay, deadline misses,
and the highest user miss ratios}. The \emph{delay cost}
$C_t^{\rm del}$ is the mean normalized delay $L/D_i$ of the URLLC packets
that complete on time in period $t$, and is zero if no packet completes on time. 

The deadline-miss penalty $[V_t-\epsilon_V]^+$ applies only to the portion of the URLLC deadline-miss ratio $V_t$ that exceeds its target $\epsilon_V$. 

The \emph{failure-tail cost} $C_t^{\rm fail}$ is the CVaR of the user miss
ratios $V_{i,t}$ for URLLC users with at least one packet becoming due in period $t$, and is zero if no user has one. Minimizing $C_t^{\rm fail}$ penalizes the highest user miss ratios. 

\paragraph{eMBB terms}
The eMBB terms are designed to \emph{maximize aggregate throughput while
minimizing severe rate shortfalls across time and users}. With a reference rate $B_{\rm ref}$, the normalized eMBB throughput is $T_t=\sum_j B_{j,t}/(|\mathcal U^{\E}|B_{\rm ref})$.
Throughput enters the cost with a negative sign, so minimizing $-T_t$ maximizes aggregate eMBB throughput. To protect individual eMBB users, we also penalize rates below $B_{\min}$. For user $j$, the squared slot deficit is $\big([(B_{\min}-B_{j,k})/B_{\min}]^+\big)^2$, and the interval deficit is defined in the same way using $B_{j,m}$. Squaring the deficit gives larger penalties to more severe shortfalls.

The \emph{time-tail cost} $C_m^{\rm time}$ is the CVaR of the slot deficits
of all eMBB users in interval $m$. Minimizing $C_m^{\rm time}$ reduces the worst rate shortfalls that occur in particular slots. 

The \emph{user-tail cost} $C_m^{\rm user}$ is the CVaR of the interval
deficits across eMBB users. Minimizing $C_m^{\rm user}$ protects the worst-served users in each interval. Let $C_t^{\rm time}$ denote the period average of $C_m^{\rm time}$.
Layer~1 forms $C_t^{\rm user}$ by averaging the largest $k_{\rm w}$
interval deficits and then averaging over the period, where $k_{\rm w}$
is a configured number of users.

\paragraph{Correction terms}
The correction magnitude between the nominal allocation $\bm a_m^{\rm nom}=(\bm x_m^{\rm nom},\bm p_m^{\rm nom})$ and the corrected allocation $\bm a_m^{\rm cor}=(\bm x_m^{\rm cor},\bm p_m^{\rm cor})$ is
\begin{equation}
\Delta a_m
=
\frac{
\left\|
\bm{x}_m^{\rm cor}-\bm{x}_m^{\rm nom}
\right\|_F^2
}{N}
+
\frac{
\left\|
\bm{p}_m^{\rm cor}-\bm{p}_m^{\rm nom}
\right\|_2^2
}{
N\left(P_{\max}^{\rm PRB}\right)^2
},
\label{eq:allocation_correction}
\end{equation}
where \(\|\cdot\|_F\) and \(\|\cdot\|_2\) denote the Frobenius and Euclidean norms, respectively.

The interval correction cost is
\begin{equation}
C_m^{\rm cor}
=
\kappa_1 \Delta a_m
+
\kappa_2 n_m^{\rm emg}
+
\kappa_3 \mathbf{1}\{n_m^{\rm emg}>0\},
\label{eq:correction_cost}
\end{equation}
where $\mathbf{1}\{\cdot\}$ is the indicator function and
$\kappa_1,\kappa_2,\kappa_3$ are positive weights. The three terms penalize the correction magnitude, the number of emergency PRB transfers, and the occurrence of an emergency transfer, respectively. The period correction cost $C_t^{\rm cor}$ is
the mean of $C_m^{\rm cor}$ over the intervals in period $t$.

To discourage abrupt changes in slice budgets between consecutive periods, we also include
\[
\Delta b_t
=
\left[
\frac{N_t^{\mathrm{U}}-N_{t-1}^{\mathrm{U}}}{N}
\right]^2
+
\left[
\frac{P_t^{\mathrm{U}}-P_{t-1}^{\mathrm{U}}}
{P_{\max}^{\rm cell}}
\right]^2.
\]

Combining these objectives with positive weights $w_1,\ldots,w_8$, the
period cost is
\begin{equation}
\begin{aligned}
C_t
={}&
w_1 C_t^{\rm del}
+w_2 [V_t-\epsilon_V]^+
-w_3 T_t
+w_4 C_t^{\rm time}
\\
&+
w_5 C_t^{\rm user}
+w_6 C_t^{\rm fail}
+w_7 C_t^{\rm cor}
+w_8 \Delta b_t .
\end{aligned}
\label{eq:period_cost}
\end{equation}

The objective is to minimize the long-run average cost $\bar C(\pi)$, balancing URLLC reliability and latency against eMBB throughput and fairness, while penalizing corrections and budget changes.
\section{Risk-Aware Problem Formulation}\label{sec:problem}

We formulate the joint slice budgeting and user allocation problem under traffic-demand and CSI uncertainty as follows:
\begin{subequations}\label{eq:master}
\begin{align}
 \min_{\pi}\quad & \bar C(\pi) \label{eq:p_obj}\\
 \text{s.t.}\quad
 & \bar V(\pi)\le\epsilon_V,\quad \bar B_j(\pi)\ge B_{\min},\ \forall j, \label{eq:p_targets}\\
 & N_t^{\U}+N_t^{\E}=N,\quad P_t^{\U}+P_t^{\E}=P^{\rm cell}_{\max},\ \forall t, \label{eq:p_budget_sum}\\
 & N_t^{\rho}\in[N_{\min}^{\rho},N_{\max}^{\rho}],\, P_t^{\rho}\in[P_{\min}^{\rho},P_{\max}^{\rho}],\ \forall\rho,t, \label{eq:p_budget_set}\\
 & \textstyle\sum_{u}x_{u,r,k}\le1,\quad x_{u,r,k}\in\{0,1\},\ \forall u,r,k, \label{eq:p_own}\\
 & 0\le p_{r,k}\le P^{\rm PRB}_{\max}\textstyle\sum_{u}x_{u,r,k},\ \forall r,k, \label{eq:p_prbpow}\\
 & \begin{aligned}
 &\textstyle\sum_{i,r}x^{\rm nom}_{i,r,m}=N_t^{\U},\ \forall m,\\
 &\textstyle\sum_{i,r}x^{\rm cor}_{i,r,m}=N_t^{\U}+n_m^{\rm emg}-n_m^{\rm rel},\ \forall m,
 \end{aligned} \label{eq:p_count}\\
 & \textstyle\sum_{r\in\mathcal N_m^{\U}}p^{\rm nom}_{r,m}\le P_t^{\U},\ \sum_{r\notin\mathcal N_m^{\U}}p^{\rm nom}_{r,m}\le P_t^{\E},\ \forall m, \label{eq:p_nom_power}\\
 & 0\le n_m^{\rm emg}\le\min\{N_{\rm emg},N_t^{\E}-N_{\rm fl}^{\E}\},\ \forall m, \label{eq:p_emg}\\
 & 0\le n_m^{\rm rel}\le\min\{N_{\rm rel},N_t^{\U}-N_{\rm fl}^{\U}\},\ \forall m, \label{eq:p_rel}\\
 & n_m^{\rm emg}n_m^{\rm rel}=0,\quad\forall m, \label{eq:p_direction}\\
 & \begin{aligned}
 &p_k^{\U}\le P_t^{\U}+\Delta P_m,\quad p_k^{\E}\le P_t^{\E}-\Delta P_m,\ \forall k,\\
 &\Delta P_m\in[-P_{\rm rel},P_{\rm emg}]\cap[-P_t^{\U},P_t^{\E}],\ \forall m,
 \end{aligned} \label{eq:p_power}\\
 & \Pr\bigl\{R_{i,k_m}\ge R_{i,m}^{\rm req}(W_{i,m})\bigr\}\ge1-\epsilon_i,\ \forall i,m, \label{eq:chance}\\
 & \bm a_k=\bm a_m^{\rm cor},\ \forall k\in\mathcal K_m,\ \forall m. \label{eq:p_hold}
\end{align}
\end{subequations}

The policy $\pi$ selects $\bm b_t$, $\bm a_m^{\rm nom}$,
$\bm a_m^{\rm cor}$, and the auxiliary grant change $\Delta P_m$
at their respective boundaries. Both allocation stages satisfy
\eqref{eq:ownership}, and corrections satisfy the transfer limits~\eqref{eq:transfer} and~\eqref{eq:power}. The objective function in~\eqref{eq:p_obj} minimizes the long-run average of the period cost~\eqref{eq:period_cost}. Constraint~\eqref{eq:p_targets} imposes the long-run service targets. Constraints~\eqref{eq:p_budget_sum}--\eqref{eq:p_power} and~\eqref{eq:p_hold} enforce the budget, assignment, transfer, power, and fixed-allocation requirements. Constraint~\eqref{eq:chance} imposes the per-user probabilistic URLLC service requirement.

Problem~\eqref{eq:master} is a stochastic mixed-integer optimization problem with a chance constraint. Its complexity arises from the coupling of decisions across two timescales and two slices, the binary assignment variables, the continuous per-PRB powers, and uncertain arrivals and channels. We therefore address it through a deterministic service test that, under stated uncertainty conditions, is sufficient for~\eqref{eq:chance} when it is satisfied at every boundary, together with a hierarchical learning framework and a safety layer that preserves the resource constraints.

\section{Uncertainty Bounds and Problem Reformulation}\label{sec:protection}

A probabilistic lower bound on the achievable rate, obtained from the predicted CSI and a Chernoff bound on the CSI prediction error, is combined with a conformal demand bound to obtain a deterministic service test for the marginal chance constraint in~\eqref{eq:chance}.

\subsection{Chernoff-Bound-Based Protected Rate}\label{subsec:protected}

The rate in \eqref{eq:rates} depends on the
realized channel magnitude $|g_{u,r,k}|$, which makes the chance
constraint~\eqref{eq:chance} difficult to handle directly. To derive
a tractable sufficient condition, we lower-bound $|g_{u,r,k}|$ with a
prescribed probability. For $|\widehat g_{u,r,k}|>0$, let
$Z_{u,r,k}=\operatorname{Re}\{\Delta g_{u,r,k}\widehat g_{u,r,k}^*\}/|\widehat g_{u,r,k}|$
denote the component of the CSI prediction error in the direction of the predicted CSI. Conditional on $\mathcal F_k$, it is
zero-mean Gaussian with standard
deviation $c_{\rm CSI}|\widehat g_{u,r,k}|$. Since
$g_{u,r,k}=\widehat g_{u,r,k}-\Delta g_{u,r,k}$ and the magnitude of $g_{u,r,k}$ is at least its component along
$\widehat g_{u,r,k}$, $|g_{u,r,k}|\ge|\widehat g_{u,r,k}|-Z_{u,r,k}$.
By the Gaussian Chernoff bound~\cite{wainwright2019}, a zero-mean
Gaussian random variable $Z$ with standard deviation $\sigma$ satisfies
$\Pr\{Z\ge z\sigma\}\le\exp(-z^2/2)$ for any $z\ge0$. Applying it to
$Z_{u,r,k}$ gives
$\Pr\{|g_{u,r,k}|<(1-zc_{\rm CSI})|\widehat g_{u,r,k}|\mid\mathcal F_k\}\le\exp(-z^2/2)$.
Splitting the CSI risk $\epsilon_{\rm CSI}$ equally over the at most
$N_{u,t}^{\rm own}$ PRBs that user $u$ can own in period $t$, i.e.,
$\min\{N,N_t^{\U}+N_{\rm emg}\}$ for a URLLC user and $N$ for an eMBB
user, and applying the union bound, we define the protected channel magnitude as
\begin{equation}
 \underline g_{u,r,k}=
 \Big[1-c_{\rm CSI}\sqrt{2\ln(N_{u,t}^{\rm own}/\epsilon_{\rm CSI})}\Big]^+
 |\widehat g_{u,r,k}|.
 \label{eq:protected_channel}
\end{equation}
We call this back-off applied to the magnitude of the predicted CSI the \emph{CSI
margin}. Conditional on $\mathcal F_k$, with probability at
least $1-\epsilon_{\rm CSI}$, $\underline g_{u,r,k}\le|g_{u,r,k}|$ on
all PRBs of user $u$. A zero prediction gives a zero protected channel magnitude, which satisfies the bound trivially. Replacing
$|g_{u,r,k}|$ with $\underline g_{u,r,k}$ in \eqref{eq:sinr} gives the
protected SNR $\underline\gamma_{u,r,k}$ under the selected power, and the protected rate is given
by
\begin{equation}
 \underline R_{u,k}=\frac{N_{\rm RE}}{\tau}\sum_{r\in\mathcal N}x_{u,r,k}\,s(\underline\gamma_{u,r,k}).
 \label{eq:protected_rate}
\end{equation}
Since $s(\cdot)$ is nondecreasing, $\underline R_{u,k}\le R_{u,k}$
with probability at least $1-\epsilon_{\rm CSI}$.

\subsection{Conformal Demand Margin}\label{subsec:margin}

The demand prediction $\widehat A_{i,k_m}$ does not, by itself, quantify the risk of underestimating demand. We therefore use conformal
prediction~\cite{angelopoulos2023} to derive a demand bound with a target coverage probability. After interval $m$, the
prediction-error score is computed as
\begin{equation}
 e_m=\max_{i\in\mathcal U^{\U},\,k\in\mathcal K_m}
 \big[A_{i,k}-\widehat A_{i,k_m}\big]^+,
 \label{eq:score}
\end{equation}
which records the largest underestimate over all URLLC users and
slots of the interval. For the demand risk $\delta$, the
\emph{demand margin} $q_m$ is the finite-sample-corrected $(1-\delta)$ quantile of the most
recent completed-interval scores, up to a fixed window length, computed as
in~\cite{angelopoulos2023}, and an initial margin is used while calibration scores accumulate. 
The \emph{conformal demand bound} of URLLC user $i$ is
then given by
\begin{equation}
 \widetilde A_{i,k_m}=\widehat A_{i,k_m}+q_m.
 \label{eq:demand_margin}
\end{equation}

We refer to
\begin{equation}
 \Pr\{e_m\le q_m\}\ge1-\delta
 \label{eq:coverage}
\end{equation}
as the \emph{demand-coverage condition}. The conformal quantile uses the calibration scores of completed intervals~\cite{angelopoulos2023}, and the demand-coverage condition can be assessed through the empirical coverage of the demand bounds. When~\eqref{eq:coverage} holds, the bounds $\widetilde A_{i,k_m}$ cover the arrivals of every URLLC user in every slot of the interval with probability at least $1-\delta$.

\subsection{Deterministic Service Test}\label{subsec:test}

The bits to serve at boundary $k_m$ are bounded by $\widetilde W_{i,m}=Q_{i,k_m}+\widetilde A_{i,k_m}$, and the
deterministic service test of the corrected allocation is given by
\begin{equation}
 \underline R_{i,k_m}(\bm a_m^{\rm cor})\ge R_{i,m}^{\rm req}(\widetilde W_{i,m}),\qquad i\in\mathcal U^{\U}.
 \label{eq:service_test}
\end{equation} Under the CSI prediction-error model and the demand-coverage condition~\eqref{eq:coverage}, the service test is a sufficient condition for the marginal probabilistic service requirement~\eqref{eq:chance} when it is satisfied at every boundary. The event
$\underline R_{i,k_m}\le R_{i,k_m}$ has probability at least
$1-\epsilon_{\rm CSI}$ by the construction of the CSI margin~\eqref{eq:protected_channel}, and the event
$e_m\le q_m$ has probability at least $1-\delta$ by
\eqref{eq:coverage}. In the second event,
$A_{i,k_m}\le\widetilde A_{i,k_m}$, so $W_{i,m}\le\widetilde W_{i,m}$, and
the required rate \eqref{eq:requirement} is nondecreasing in $W$. Hence,
when both events occur, an allocation that satisfies
\eqref{eq:service_test} gives
$R_{i,k_m}\ge\underline R_{i,k_m}\ge R_{i,m}^{\rm req}(\widetilde W_{i,m})
\ge R_{i,m}^{\rm req}(W_{i,m})$. By the union bound, both events occur
with probability at least $1-\epsilon_{\rm CSI}-\delta$. Thus, the probability that the test passes and the required rate is not met is at most $\epsilon_{\rm CSI}+\delta$. If the allocation satisfies the test at every boundary, the marginal service requirement~\eqref{eq:chance} holds with $\epsilon_i=\epsilon_{\rm CSI}+\delta$. 

Replacing~\eqref{eq:chance} by the sufficient test~\eqref{eq:service_test}
gives the following conservative stochastic formulation:
\begin{equation}
 \min_{\pi}\ \bar C(\pi)\quad\text{s.t.}\quad
 \eqref{eq:p_targets}\text{--}\eqref{eq:p_power},\ \eqref{eq:service_test},\ \eqref{eq:p_hold}.
 \label{eq:det_problem}
\end{equation}
The test applies to the backlog and the arrivals at the assignment boundary. Arrivals in later slots of the interval are served by the fixed allocation, and any remaining backlog enters the test at the next boundary.

\section{Risk-Aware Safe Hierarchical Learning Framework}\label{sec:hierarchy}

Slice budgets and user assignments belong to different O-RAN control loops. A single policy would set the URLLC PRB count implicitly through $|\mathcal U|^{N}$ assignments and trade URLLC deadlines directly against eMBB rates through a shared reward. Separate policies address each service through discounted training costs, while a deterministic check tests the per-user service requirement before execution. Layer~1 determines the resources available to Layers~2A, 2B, and~3. The correction cost $C_t^{\rm cor}$ enters its state and the period cost~\eqref{eq:period_cost}, penalizing budgets that repeatedly require correction.

Layer~1 selects $\bm b_t$ once per orchestration period using reward $-C_t$, with decoding enforcing~\eqref{eq:p_budget_sum}--\eqref{eq:p_budget_set}. At each assignment boundary, Layer~2A selects the URLLC PRBs, users, and powers subject to~\eqref{eq:p_own}--\eqref{eq:p_prbpow}, the nominal count in~\eqref{eq:p_count}, and~\eqref{eq:p_nom_power}. Layer~2B assigns the remaining PRBs and powers to eMBB subject to~\eqref{eq:p_own}--\eqref{eq:p_prbpow} and~\eqref{eq:p_nom_power}. The long-run targets~\eqref{eq:p_targets} enter training through service penalties and the adaptive multiplier and are checked in evaluation. Layer~3 seeks a correction satisfying~\eqref{eq:service_test}, \eqref{eq:p_own}--\eqref{eq:p_count}, and~\eqref{eq:p_emg}--\eqref{eq:p_power}, with best-effort fallback if the test cannot be met. The corrected allocation remains fixed as required by~\eqref{eq:p_hold}.

\subsection{Layer 1: Slice Budget Policy}\label{subsec:layer1}

\emph{a) State:} Layer~1 observes summary statistics of both slices (Table~\ref{tab:layers}). For URLLC, these include the backlogs, the conformal demand bounds, the queue-based delay $Q_i/\underline R_i+\phi_i$, and packet urgency, i.e., the age of the oldest queued packet, its time to deadline, and the due bits $W_{i,m}^{\rm due}(\ell)$ in~\eqref{eq:requirement}. For eMBB, they include the backlogs, demand predictions, delivered rates, and the CVaR of recent squared rate deficits below $B_{\min}$. Both slices also contribute features of the protected channel magnitudes. In addition, Layer~1 observes three quantities from the previous period: the budget $\bm b_{t-1}$, the correction cost $C_{t-1}^{\rm cor}$, and the delay-violation frequency, i.e., the fraction of URLLC user--slot pairs whose backlog-to-rate ratio exceeded the deadline or whose bits overflowed the buffer.

\emph{b) Action:} The action $(\theta_t,\alpha_t)$ is selected from a finite set in $[0,1]^2$. Here, $\theta_t$ sets the URLLC share of the PRBs, and $\alpha_t$ sets the URLLC power within its admissible range. The URLLC PRB budget is $N_t^{\U}=\lceil\theta_tN\rceil$, clipped to the range allowed by the bounds of both slices in~\eqref{eq:budget_set}. The URLLC power budget interpolates between a lower limit $P_{\rm lo}$ and an upper limit $P_{\rm hi}$:
\begin{equation}
 P_t^{\U}=P_{\rm lo}(N_t^{\U})+
 \alpha_t\big[P_{\rm hi}(N_t^{\U})-P_{\rm lo}(N_t^{\U})\big].
 \label{eq:budget_decode}
\end{equation}
The eMBB slice receives the remaining PRBs and power. The limits depend on the largest number of PRBs that each slice can hold after a Layer~3 transfer, $\widehat N_t^{\U}=N_t^{\U}+\min\{N_{\rm emg},[N_t^{\E}-N_{\rm fl}^{\E}]^+\}$ and $\widehat N_t^{\E}=N_t^{\E}+\min\{N_{\rm rel},[N_t^{\U}-N_{\rm fl}^{\U}]^+\}$:
\begin{equation*}
\begin{aligned}
P_{\rm lo}(N_t^{\U})&=\max\{P_{\min}^{\U},\,P_{\max}^{\rm cell}-P_{\max}^{\E},\,P_{\max}^{\rm cell}-\widehat N_t^{\E}P_{\rm ref}^{\rm PRB}\},\\
P_{\rm hi}(N_t^{\U})&=\min\{P_{\max}^{\U},\,P_{\max}^{\rm cell}-P_{\min}^{\E},\,\widehat N_t^{\U}P_{\rm ref}^{\rm PRB}\}.
\end{aligned}
\end{equation*}
The first two terms of each limit keep the power budgets of both slices within the bounds in~\eqref{eq:budget_set}. The third term caps the power of each slice at what its largest PRB count can use at the reference per-PRB power $P_{\rm ref}^{\rm PRB}$. Since $P_{\rm ref}^{\rm PRB}<P_{\max}^{\rm PRB}$, Layers~2A and~2B can still raise individual PRB powers above this reference. With $P_{\rm lo}\le P_{\rm hi}$, every action yields a budget that satisfies~\eqref{eq:p_budget_sum} and~\eqref{eq:p_budget_set}.

\emph{c) Reward:} The reward is $-C_t$, with $C_t$ given in~\eqref{eq:period_cost}. Since $C_t$ includes the correction cost, Layer~1 is also penalized for budgets that frequently require Layer~3 corrections.
\subsection{Layers 2A and 2B: PRB and Power Allocation Policies}\label{subsec:layer2}

Separate policies allow each service to use its own reward and reduce the action space of each agent. Layer~2A acts first to prioritize the shorter URLLC deadlines.

\emph{a) State:} Layer~2A uses URLLC queue, deadline, demand and channel information. Layer~2B uses eMBB traffic, rate and channel information, together with the remaining PRBs. Both include the slice budgets and previous assignments, namely the PRB owners in the preceding interval (Table~\ref{tab:layers}).

Rolling eMBB rates average recent slots. Smoothed rates use
a running average initially, followed by exponentially weighted updates. Rate deficits compare these rates with $B_{\min}$ and with the mean eMBB user rate. The candidate protected rates evaluate~\eqref{eq:protected_rate} at powers obtained by scaling the slice's average power per PRB by the allowed weights in $\Omega$, subject to $P_{\max}^{\rm PRB}$.

\emph{b) Action:} Layer~2A selects the $N_t^{\U}$ PRBs reserved for URLLC and assigns each to a URLLC user with its transmit power. Layer~2B assigns the remaining PRBs and their transmit powers to eMBB users. For each URLLC PRB, it also selects an eMBB \emph{shadow owner}, which receives the PRB only if Layer~3 releases it.

Users are assigned in PRB order, with each assignment conditioned on the preceding allocations. For URLLC, the policy also considers the protected service already allocated to each user relative to its demand.

After user assignment, each policy selects a positive power weight $\omega_{r,m}\in\Omega$ for each PRB of its slice. Let $\mathcal N_m^{\E}$ denote the PRBs remaining after the URLLC selection. The nominal powers are
\begin{equation}
 p^{\rm nom}_{r,m}=\min\Big\{P^{\rm PRB}_{\max},\
 \frac{\omega_{r,m}P_t^{\rho}}{\sum_{r'\in\mathcal N_m^{\rho}}\omega_{r',m}}\Big\},\quad r\in\mathcal N_m^{\rho}.
 \label{eq:power_decode}
\end{equation}
The selected weights distribute the slice power budget among its PRBs subject to the per-PRB limit. Hence, every weight selection satisfies \eqref{eq:p_prbpow} and \eqref{eq:p_nom_power}. Unassigned PRBs have zero power, and power removed by clipping remains unused.

\emph{c) Reward:} Layer~2A uses reward $-C_m^{\rm 2A}-\mu V_m$, where $V_m$ is the interval miss ratio, $\mu\ge0$ is an adaptive penalty multiplier, and
\begin{equation}
 C_m^{\rm 2A}=\lambda_1C_m^{\rm Q}+\lambda_2C_m^{\rm urg}+\lambda_3C_m^{\rm del}
 +\lambda_4C_m^{\rm lat}+\lambda_5C_m^{\rm fail}.
 \label{eq:l2a_cost}
\end{equation}
This cost combines backlog, urgency, delay, and failure penalties with positive weights, providing interval-level feedback for the URLLC objectives in~\eqref{eq:period_cost}. The backlog cost $C_m^{\rm Q}$ is the mean backlog after service across URLLC users and slots, normalized by $Q_i^{\rm safe}$. The urgency cost $C_m^{\rm urg}$ increases the penalty on interval-end backlog as the queue-to-rate delay estimate, including processing delay, increases relative to the deadline. The delay costs $C_m^{\rm del}$ and $C_m^{\rm lat}$ are the mean and CVaR of normalized on-time packet delays. The failure cost $C_m^{\rm fail}$ is the CVaR of per-user interval miss ratios.

The additional penalty $\mu V_m$ discourages deadline misses, linking the reward to the reliability target in~\eqref{eq:p_targets}. Here, $V_m$ is the fraction of failed packets among those successfully completing or first failing in interval $m$, and is zero when no such event occurs. In contrast, $V_t$ groups packets by when they fall due. The multiplier is adjusted during policy updates with step size $\eta$ and cap $\mu_{\max}$ using the auxiliary training threshold $\epsilon_{\rm 2A}$, while $\epsilon_V$ remains the long-run reliability target.

Layer~2B uses reward $-C_m^{\rm 2B}$, where
\begin{equation}
 C_m^{\rm 2B}=-\lambda_6\Psi_m+\lambda_7C_m^{\rm time}+\lambda_8C_m^{\rm user}.
 \label{eq:l2b_cost}
\end{equation}
With positive weights, this cost balances proportional-fair eMBB utility against rate shortfalls across slots and users, reflecting the eMBB objectives of~\eqref{eq:period_cost}. The utility $\Psi_m$ is the mean of $\ln(1+B_{j,m}/B_{\rm off})$ across eMBB users, divided by $\ln(1+B_{\rm ref}/B_{\rm off})$, where $B_{\rm off}$ is a small rate offset and $B_{\rm ref}$ is the reference rate. The costs $C_m^{\rm time}$ and $C_m^{\rm user}$ penalize the slot and user rate shortfalls defined in the cost model. Both policies maximize the discounted returns associated with their interval costs as training objectives related to the long-run average~\eqref{eq:p_obj}.

\subsection{Layer 3: Uncertainty-Aware Safety Layer}\label{subsec:layer3}

Layer~3 is a deterministic algorithm that checks and, when necessary, corrects the nominal
allocation before execution, following the safety-layer approach
in~\cite{dalal2018}. At each assignment boundary, it verifies constraints~\eqref{eq:ownership} and~\eqref{eq:power}, evaluates the service
test~\eqref{eq:service_test} for every URLLC user using the protected channel magnitudes and conformal demand bounds, and handles the following 
cases.

\emph{Deficit:} If at least one URLLC user fails the service test, Layer~3 performs
the \emph{minimum emergency transfer}, which minimizes the number
of eMBB PRBs moved to URLLC users subject to the service test:
\begin{equation}
 \begin{aligned}
 \bm a_m^{\rm cor}=\arg\min_{\bm a}\ &n_m^{\rm emg}(\bm a)\\
 \text{s.t.}\ &\underline R_{i,k_m}(\bm a)\ge R_{i,m}^{\rm req}(\widetilde W_{i,m}),
 \quad i\in\mathcal U^{\U},\\
 &\eqref{eq:ownership},\ \eqref{eq:transfer},\ \eqref{eq:power},
 \end{aligned}
 \label{eq:layer3}
\end{equation}
where $n_m^{\rm emg}(\bm a)$ counts the eMBB PRBs of $\bm a_m^{\rm nom}$
that are moved to URLLC users in $\bm a$, and $\bm a$ equals
$\bm a_m^{\rm nom}$ on all other PRBs. Since $N_{\rm emg}$ is small, the algorithm enumerates admissible PRB subsets and their URLLC users to identify a minimum-cardinality feasible correction. Among corrections requiring the same minimum number of transfers, it takes PRBs first from the eMBB users with the lowest required rate over the rest of the period, defined as the remaining bits needed to meet $B_{\min}$ divided by the remaining period duration, to limit the impact on underserved users. A transferred PRB retains its nominal power, and only corrected URLLC power above $P_t^{\U}$ counts against $P_{\rm emg}$ in~\eqref{eq:power}. If no
allocation within the caps satisfies the test, Layer~3 transfers, up
to the cap, the PRBs that most reduce the total URLLC rate deficit. Execution retains residual deficit $\sum_i[R_{i,m}^{\rm req}(\widetilde W_{i,m})-\underline R_{i,k_m}(\bm a_m^{\rm cor})]^+$, and remaining queued bits enter the next boundary test.

\emph{Surplus:} If every URLLC user passes the service test but an eMBB user remains below its rate target, i.e., the smaller of $B_{\min}$ and its offered rate over a sliding window of $K$ slots, Layer~3 releases up to $N_{\rm rel}$ URLLC PRBs to eMBB, starting with those that contribute least to the protected URLLC rates. A candidate released PRB is assigned to its shadow owner, selected by the Layer~2B conditional owner policy, provided that user has demand and its protected rate over the window remains below its rate target. The owners of all other PRBs remain fixed, as do the powers of PRBs that remain assigned to URLLC. Released PRBs join the eMBB set with unit weight, primary eMBB PRBs keep their selected weights, and the eMBB budget is redistributed over the enlarged set using~\eqref{eq:power_decode}. A release is accepted only if every URLLC user still passes the service test and the protected eMBB sum rate does not decrease.

\emph{No correction:} If neither condition warrants an admissible transfer, Layer~3 keeps the nominal allocation.

Table~\ref{tab:layers} summarizes the state, action, and reward of each layer. 

\begin{table*}[t]
\centering\caption{State, action, and reward by layer. Layer~1 acts once per orchestration period $t$; Layers~2A, 2B, and~3 act once per assignment interval $m$. Queues, traffic predictions, and CSI are shown without time indices.}
\label{tab:layers}\footnotesize\renewcommand{\arraystretch}{1.0}
\setlength{\heavyrulewidth}{0.6pt}
\setlength{\lightrulewidth}{0.3pt}
\setlength{\tabcolsep}{4pt}
\begin{tabular*}{\textwidth}{@{\extracolsep{\fill}}L{.035\textwidth}L{.50\textwidth}L{.30\textwidth}L{.105\textwidth}@{}}
\toprule
\textbf{Layer} & \textbf{State} & \textbf{Action} & \textbf{Reward} \\
\midrule
\textbf{1} & URLLC: $Q_i$, $\widetilde A_i$, delay $Q_i/\underline R_i+\phi_i$, urgency, $W_{i,m}^{\rm due}(\ell)$; eMBB: $Q_j$, $\widehat A_j$, $B_{j,k}$, $\cvar$ of past deficits; features of $\underline g_{u,r}$; $\bm b_{t-1}$, $C_{t-1}^{\rm cor}$, delay-violation frequency
& $(\theta_t,\alpha_t)\mapsto\bm b_t=(N_t^{\U},P_t^{\U})$ via~\eqref{eq:budget_decode}, under \eqref{eq:p_budget_sum}--\eqref{eq:p_budget_set}
& $-C_t$ \\
\addlinespace[2pt]
\textbf{2A} & $Q_i$, packet ages, times to deadline, $W_{i,m}^{\rm due}(\ell)$; features of $\widehat g_{i,r},\underline g_{i,r}$; $\widehat A_i$, $\widetilde A_i$; $\underline R_i$ at $\omega\in\Omega$; $\bm b_t$; $\bm x_{m-1}^{\rm cor}$
& $\mathcal N_m^{\U}$, $x_{i,r,m}$, $\omega_{r,m}\mapsto p_{r,m}^{\rm nom}$ via~\eqref{eq:power_decode}, under \eqref{eq:p_own}--\eqref{eq:p_nom_power}
& $-C_m^{\rm 2A}-\mu V_m$ \\
\addlinespace[2pt]
\textbf{2B} & $Q_j$, $\widehat A_j$; features of $\widehat g_{j,r},\underline g_{j,r}$; rolling and smoothed $B_{j,k}$ and deficits; $\underline R_j$ at $\omega\in\Omega$; $\bm b_t$; $\mathcal N_m^{\E}$; $\bm x_{m-1}^{\rm cor}$
& $x_{j,r,m}$, $\omega_{r,m}\mapsto p_{r,m}^{\rm nom}$ on $\mathcal N_m^{\E}$; shadow owners on $\mathcal N_m^{\U}$, under \eqref{eq:p_own}, \eqref{eq:p_prbpow}, \eqref{eq:p_nom_power}
& $-C_m^{\rm 2B}$ \\
\addlinespace[2pt]
\textbf{3} & $\bm a_m^{\rm nom}$, eMBB $\omega_{r,m}$, shadow owners; $\bm b_t$; $Q_u$, $W_{i,m}^{\rm due}(\ell)$, $\widetilde A_i$, $\underline g_{u,r}$; $K$-slot window rates (projected protected, offered capped at $B_{\min}$)
& $n_m^{\rm emg}$ at fixed powers or $n_m^{\rm rel}$ with eMBB powers re-decoded, $\Delta P_m$, giving $\bm a_m^{\rm cor}$ under \eqref{eq:p_own}--\eqref{eq:p_count}, \eqref{eq:p_emg}--\eqref{eq:p_power}
& None \\
\bottomrule
\end{tabular*}
\end{table*}

\subsection{Policy Training}\label{subsec:training}

Layers~1, 2A, and~2B are trained as separate PPO agents with generalized advantage estimation~\cite{schulman2017ppo}, each with an actor over its masked discrete actions and a critic. PPO is used because its clipped updates discourage large policy changes, which is useful when optimizing one timescale while the other remains fixed. The discount factor of Layer~1 per period is the $M$-th power of that of Layers~2A and~2B per interval, so all layers discount time at the same rate. Since training both timescales at once would make each learn against policies that are still changing, training alternates between them in four phases. In phases~1 and~3, Layers~2A and~2B are trained under varied admissible budgets and under budgets selected by the fixed Layer~1 policy, respectively. In phases~2 and~4, Layer~1 is trained with the assignment policies fixed. The power weights remain at unit value in phases~1 and~2, and Layers~2A and~2B learn them from phase~3 onward. During testing, the network parameters remain fixed, while the queues, predictions, prediction-error scores, and budget histories continue to evolve.

\subsection{Computational Complexity}\label{subsec:runtime}

Layer~1 evaluates a small network on a fixed-size observation once per orchestration period, so its cost is $\mathcal O(1)$ in the numbers of PRBs and users. Layers~2A and~2B apply one shared scorer, with $n_{\rm in}$ inputs per pair and two hidden layers of $H$ units, to each PRB--user pair, so their costs, $\mathcal O\big(|\mathcal U^{\U}|(N+N_t^{\U})(n_{\rm in}H+H^2)\big)$ and $\mathcal O\big(|\mathcal U^{\E}|N(n_{\rm in}H+H^2)\big)$, grow linearly with the numbers of users and PRBs. The additional term of Layer~2A arises from conditioning the owners of its $N_t^{\U}$ PRBs on earlier owner choices. The exact-cardinality subset sampler adds $\mathcal O(NN_t^{\U})$ work. The power head adds $\mathcal O\big(N(n_{\rm pw}H_{\rm p}+H_{\rm p}^2)\big)$ per layer, with $n_{\rm pw}$ inputs and two hidden layers of $H_{\rm p}$ units. Layer~3 checks each candidate transfer against the required rates of the $|\mathcal U^{\U}|$ URLLC users, so its cost is $\mathcal O\big((N_t^{\E})^{N_{\rm emg}}|\mathcal U^{\U}|^{N_{\rm emg}+1}\big)$, which grows combinatorially with the transfer cap.

\section{Performance Evaluation}\label{sec:eval}

This section evaluates the proposed framework against five baselines, comprising three learning-based methods and two fixed schedulers with static slice budgets, under a common cell, traffic, and channel model.

\subsection{Evaluation Setup}\label{subsec:setup}

Table~\ref{tab:setup} summarizes the evaluation and training parameters, and the traffic parameters define the nominal load. The 4~W per-PRB power limit allows power to be concentrated on selected PRBs, while the reference $P_{\rm ref}^{\rm PRB}$ is used only to decode the Layer~1 budget range. The 10~ms assignment interval and the 1~s orchestration period correspond to the shortest near-RT and non-RT control loops considered in~\cite{polese2023oran}. A pretrained LSTM is used for channel-gain prediction. Users are initially uniformly distributed between 50 and 300~m from the base station and move within these bounds according to a Gauss--Markov mobility model. The value $c_{\rm CSI}=0.08$ corresponds to a normalized mean-squared error of $-18.9$~dB relative to the predicted-CSI power. Separate LSTM networks for URLLC and eMBB predict the per-user next-slot arrivals. Since the model does not include decoding errors, $\epsilon_V=0.05$ is a scheduling-level target rather than a radio-link reliability target. Each test run evaluates one trained model on one realization of traffic, channels, and mobility, with the same realizations used across methods and excluded from training.

\begin{table}[t]
\centering\caption{Evaluation and training settings.}\label{tab:setup}
\footnotesize\renewcommand{\arraystretch}{1.0}\setlength{\tabcolsep}{2pt}
\begin{tabular*}{\columnwidth}{@{\extracolsep{\fill}}L{.24\columnwidth}L{.73\columnwidth}@{}}
\toprule
\textbf{Parameter} & \textbf{Setting} \\
\midrule
Cell and carrier & 6 URLLC, 8 eMBB users; 10~MHz; $N=24$ PRBs, 360~kHz each. \\
Timescales & $\tau=0.5$~ms; $K=20$ (10~ms); $M=100$ (1~s). \\
Budget limits & $P^{\rm cell}_{\max}=40$~W; $P^{\rm PRB}_{\max}=4$~W and $P^{\rm PRB}_{\rm ref}=2$~W; $P_{\min}^{\rho}=2$~W, $P_{\max}^{\rho}=38$~W; $N_{\min}^{\rho}=2$, $N_{\max}^{\rho}=22$. \\
Channel & 50--300~m; path-loss exponent 3.5, reference loss 38~dB; shadowing 6~dB, fixed per run; fading correlation 0.995, independent across PRBs. \\
Link mapping & CQI Table~2; $P_{\rm noise}=7.2\times10^{-15}$~W; $N_{\rm RE}=150$; $10\log_{10}\Gamma=1.5$~dB. \\
Traffic & URLLC: 256-bit packets at 0.1 (calm) or 8 (burst) per slot; eMBB: 8000-bit packets at 0.5 or 2 per slot. \\
Traffic dynamics & Calm-to-burst and burst-to-calm probabilities: 0.002, 0.01 (URLLC), 0.02, 0.1 (eMBB); load factors: 0.25, 1 per 80 slots (URLLC), 0.25, 1, 2 per 2000 slots (eMBB). \\
Buffers & $Q_i^{\max}=100$~kbit; $Q_j^{\max}=50$~Mbit; $Q_i^{\rm safe}=2000$~bits. \\
Predictors & LSTMs, 32 hidden units: demand (16-slot input), Layer~1 channel feature (20-slot input). \\
Service targets & $D_i=5$~ms; $\phi_i=0.5$~ms; $\epsilon_V=0.05$; $B_{\min}=2$~Mbps; $B_{\rm ref}=4$~Mbps; $B_{\rm off}=1$~kbps. \\
Protection & $c_{\rm CSI}=0.08$; $\epsilon_{\rm CSI}=0.001$; $\delta=0.1$ ($\epsilon_i=0.101$); 2048-score window; initial margin 512~bits. \\
Transfer caps & $N_{\rm emg}=2$, $P_{\rm emg}=8$~W; $N_{\rm rel}=4$, $P_{\rm rel}=0$; $N_{\rm fl}^{\rho}=2$. \\
Layer-1 actions & $\theta_tN\in\{6,8,\ldots,16\}$; $\alpha_t\in\{0.5,0.6\}$. \\
Power weights & $\Omega=\{0.25,0.5,1,2,4\}$. \\
Policy training & Four phases (assignment, budget, assignment, budget): 480, 640, 480, 640~s; PPO learning rate 0.0003, clip 0.2, 12 epochs. \\
Miss multiplier & $\epsilon_{\rm 2A}=0.02$; initial $\mu=5$; $\eta=10$; $\mu_{\max}=100$. \\
CVaR levels and user tail & $\beta=0.95$ (time and latency), 0.75 (eMBB user), $0.\overline{6}$ (failure), $k_{\rm w}=2$ (Layer~1). \\
Weights & $(w_1,\ldots,w_8)=(1,20,3,0.5,1,2,0.2,0.05)$; $(\lambda_1,\ldots,\lambda_8)=(1,1,1,0.25,2,3,0.5,8)$; $(\kappa_1,\kappa_2,\kappa_3)=(0.25,1,1)$. \\
Networks & Two hidden layers: 32 units (Layer~1), $H=64$ (Layers~2A and~2B), power head $H_{\rm p}=32$, $n_{\rm pw}=53$ (2A), 23 (2B); $n_{\rm in}=51$ (2A), 20 (2B); interval discount 0.99. \\
Test runs & 10 training seeds; 40 common test environments (10 separate ones for ablations); 6~s, empty initial queues and score windows; fixed network parameters. \\
\bottomrule
\end{tabular*}
\end{table}

\begin{table*}[t]
\centering
\caption{Performance at nominal load. Learned methods report mean $\pm$ standard deviation over ten trained models.}
\label{tab:nominal}
\footnotesize\renewcommand{\arraystretch}{1.0}
\setlength{\heavyrulewidth}{0.6pt}
\setlength{\lightrulewidth}{0.3pt}
\setlength{\cmidrulewidth}{0.3pt}
\setlength{\tabcolsep}{3pt}
\begin{tabular*}{\textwidth}{@{\extracolsep{\fill}}l*{5}{r@{\,}c@{\,}l}@{}}
\toprule
& \multicolumn{6}{c}{\textbf{URLLC}} & \multicolumn{9}{c}{\textbf{eMBB}}\\
\cmidrule(lr){2-7}\cmidrule(l){8-16}
\textbf{Algorithm}
& \multicolumn{3}{c}{\shortstack{Miss ratio\\$V$ (\%) $\downarrow$}}
& \multicolumn{3}{c}{\shortstack{Worst-user miss ratio\\$V_{\rm worst}$ (\%) $\downarrow$}}
& \multicolumn{3}{c}{\shortstack{Aggregate throughput\\$B_{\rm sum}$ (Mbps) $\uparrow$}}
& \multicolumn{3}{c}{\shortstack{Worst-user rate\\$B_{\rm worst}$ (Mbps) $\uparrow$}}
& \multicolumn{3}{c}{\shortstack{Jain fairness\\$\Phi$ $\uparrow$}}\\
\midrule
\textbf{Proposed}
& $1.35$ & $\pm$ & $0.97$
& $4.84$ & $\pm$ & $3.54$
& $26.16$ & $\pm$ & $1.57$
& $2.52$ & $\pm$ & $0.18$
& $0.985$ & $\pm$ & $0.005$\\
\addlinespace[2pt]
Joint DRL
& $30.08$ & $\pm$ & $7.73$
& $89.68$ & $\pm$ & $9.63$
& $8.75$ & $\pm$ & $2.74$
& $0.00$ & $\pm$ & $0.00$
& $0.180$ & $\pm$ & $0.060$\\
SafeSlice
& $4.22$ & $\pm$ & $0.64$
& $8.41$ & $\pm$ & $0.60$
& $6.05$ & $\pm$ & $0.44$
& $0.48$ & $\pm$ & $0.03$
& $0.958$ & $\pm$ & $0.000$\\
Two-timescale slicing
& $7.03$ & $\pm$ & $4.42$
& $11.59$ & $\pm$ & $6.15$
& $29.83$ & $\pm$ & $5.59$
& $2.34$ & $\pm$ & $0.45$
& $0.957$ & $\pm$ & $0.001$\\
\addlinespace[3pt]
Fixed PF
& $3.73$ & &
& $7.57$ & &
& $23.92$ & &
& $1.87$ & &
& $0.956$ & &\\
Fixed floor-first PF
& $3.73$ & &
& $7.57$ & &
& $22.56$ & &
& $1.95$ & &
& $0.878$ & &\\
\bottomrule
\end{tabular*}
\end{table*}

\subsection{Baseline Algorithms}\label{subsec:baselines}

All baselines use the same single-cell setup and are trained for 2240~s at each operating point. Learning-based baselines retain their original rewards and learning rates unless stated otherwise, and are evaluated using their final trained models.

\begin{enumerate}
\item \emph{Joint DRL}~\cite{sohaib2024} was designed for URLLC puncturing across four cells. Our single-cell adaptation updates slice PRB and power budgets once per second and user allocation, idle-PRB reclaim, and eMBB power every 10~ms using channel and traffic features. It does not implement reactive per-slot puncturing. At nominal load, eMBB power is learned; in the carrier-size and user-count experiments, power is shared equally within each slice. It has no execution-time safety layer.
\item \emph{SafeSlice}~\cite{safeslice2025} selects slice bandwidth every 100~ms and corrects actions using a learned latency-cost model. We fit the model to collected state--action data, with mean queue-based delay as each slice's target, although it does not meet all source-method fidelity criteria. Since SafeSlice does not control power, we use equal slice power budgets and equal power sharing within each slice. We replace round-robin scheduling with fixed proportional fair (PF) because it caused frequent URLLC deadline misses. Within slice budgets, URLLC PRBs are assigned using the rate needed to clear queued bits by the 5~ms deadline, with an allowance for recent arrivals.
\item \emph{Two-timescale slicing} keeps the recurrent slice controller of~\cite{setayesh2022}, which updates the slice PRB and power budgets once per second, together with the share of the URLLC reservation accessible to eMBB. We replace the attention-based user schedulers with fixed PF because a backlog-proportional URLLC split used in their place led to frequent URLLC deadline misses. The controller is trained using stochastic gradient descent with a learning rate of 0.1. Between slice-budget updates, fixed PF assigns PRBs to users, with equal power sharing within each slice. There is no near-RT safety correction.
\item \emph{Fixed PF} and \emph{fixed floor-first PF} reserve half the
PRBs for URLLC ($\theta=0.5$, $\alpha=0.6$). Both assign URLLC PRBs using the rate needed to clear each user's queued bits within the 5~ms deadline, with an allowance for recent arrivals.
For eMBB, fixed PF uses PF scheduling~\cite{kelly1998}, whereas fixed floor-first PF prioritizes users with rates below $B_{\min}$. Their URLLC results are identical. Both update every 
10~ms without a safety layer.
\end{enumerate}

\subsection{Performance Metrics}\label{subsec:protocol}

We evaluate the two service targets in~\eqref{eq:service_targets}, together with the delivered eMBB throughput. The miss ratio $V$, the run estimate of the long-run packet-weighted ratio $\bar V$ in~\eqref{eq:service_targets}, divides the number of failed URLLC packets falling due during the run, including packets dropped from queues, by the number of URLLC packets falling due during the run~\cite{3gpp22261reliability}.

For a run of $n_{\rm p}$ periods,
\begin{equation}
 B_{\rm sum}=\sum_{j\in\mathcal U^{\E}}B_j^{\rm run},\qquad
 B_{\rm worst}=\frac{1}{n_{\rm p}}\sum_{t=0}^{n_{\rm p}-1}
 \min_{j\in\mathcal U^{\E}}B_{j,t},
 \label{eq:embb_metrics}
\end{equation}
where $B_j^{\rm run}=\frac{1}{n_{\rm p}}\sum_{t=0}^{n_{\rm p}-1}B_{j,t}$ is
the average rate of user $j$ over the run. All period rates use a
one-second window.

Jain's fairness index~\cite{jain1984fairness} measures the balance of delivered eMBB rates:
\begin{equation}
 \Phi=\Bigl(\textstyle\sum_{j\in\mathcal U^{\E}}B_j^{\rm run}\Bigr)^2\Big/
 \Bigl(|\mathcal U^{\E}|\textstyle\sum_{j\in\mathcal U^{\E}}(B_j^{\rm run})^2\Bigr).
 \label{eq:jain_metric}
\end{equation}
For URLLC, the worst-user miss ratio $V_{\rm worst}$, the largest miss ratio of a URLLC user over the run, captures failures that a pooled average can conceal. Joint satisfaction $J$ is the percentage of test runs that meet both criteria, $V\le\epsilon_V$ and $B_{\rm worst}\ge B_{\min}$. Since $B_{\rm worst}$ is the average of the per-period minimum user rates, it provides a conservative test of the eMBB rate target in~\eqref{eq:service_targets}. For two-timescale slicing, eight samples of its stochastic policy are aggregated within each model--environment pair for $J$ and all metrics, with $V_{\rm worst}$ computed per sample and then averaged.

\subsection{Performance Comparison at Nominal Load}\label{subsec:nominal}

Table~\ref{tab:nominal} shows that the proposed framework is the only evaluated method that meets, on average, both the 5\% URLLC deadline-miss target and the 2~Mbps worst-user eMBB rate criterion. Among the evaluated methods, it attains the lowest mean and worst-user URLLC miss ratios and the highest worst-user eMBB rate and Jain fairness, while only two-timescale slicing achieves higher aggregate eMBB throughput (29.83 vs.\ 26.16~Mbps), at a miss ratio above the URLLC target (7.03\%). Compared with the fixed schedulers, which have the lowest miss ratios among the baselines, it reduces the mean miss ratio by 64\% (1.35\% vs.\ 3.73\%) and increases aggregate throughput by up to 16\% (26.16 vs.\ 22.56~Mbps).

Joint DRL targets eMBB sum rate without incorporating queue state and concentrates resources on a few users, with low fairness ($\Phi=0.180$) and a near-zero worst-user eMBB rate. SafeSlice allocates only 63\% of the PRBs because its objective rewards unused resources, leaving a worst-user eMBB rate of 0.48~Mbps. The fixed schedulers satisfy the pooled URLLC target, but their worst-user eMBB rates remain below 2~Mbps. The proposed framework meets the deadline-miss target in 95\% of runs and the worst-user eMBB rate criterion in 86.25\%, compared with 82.5\% and at most 52.5\% for the fixed schedulers.

\begin{table*}[t]
\centering\caption{Ablation study at nominal load (mean $\pm$ standard deviation over ten trained models).}\label{tab:ablation}
\footnotesize\renewcommand{\arraystretch}{1.0}
\setlength{\heavyrulewidth}{0.6pt}
\setlength{\lightrulewidth}{0.3pt}
\setlength{\cmidrulewidth}{0.3pt}
\setlength{\tabcolsep}{3pt}
\begin{tabular*}{\textwidth}{@{\extracolsep{\fill}}l*{5}{r@{\,}c@{\,}l}@{}}
\toprule
& \multicolumn{6}{c}{\textbf{URLLC}} & \multicolumn{9}{c}{\textbf{eMBB}}\\
\cmidrule(lr){2-7}\cmidrule(l){8-16}
\textbf{Configuration} & \multicolumn{3}{c}{\shortstack{Miss ratio\\$V$ (\%) $\downarrow$}} & \multicolumn{3}{c}{\shortstack{Worst-user miss ratio\\$V_{\rm worst}$ (\%) $\downarrow$}} & \multicolumn{3}{c}{\shortstack{Aggregate throughput\\$B_{\rm sum}$ (Mbps) $\uparrow$}} & \multicolumn{3}{c}{\shortstack{Worst-user rate\\$B_{\rm worst}$ (Mbps) $\uparrow$}} & \multicolumn{3}{c}{\shortstack{Jain fairness\\$\Phi$ $\uparrow$}}\\
\midrule
\textbf{Complete framework} & $2.64$ & $\pm$ & $1.27$ & $8.65$ & $\pm$ & $4.88$ & $26.70$ & $\pm$ & $1.37$ & $2.84$ & $\pm$ & $0.18$ & $0.996$ & $\pm$ & $0.002$\\
\addlinespace[2pt]
Without eMBB CVaR & $2.12$ & $\pm$ & $0.98$ & $8.73$ & $\pm$ & $3.68$ & $24.62$ & $\pm$ & $1.75$ & $2.45$ & $\pm$ & $0.35$ & $0.988$ & $\pm$ & $0.011$\\
Without URLLC CVaR & $3.61$ & $\pm$ & $0.86$ & $9.62$ & $\pm$ & $2.67$ & $28.81$ & $\pm$ & $0.32$ & $3.01$ & $\pm$ & $0.14$ & $0.995$ & $\pm$ & $0.002$\\
Without any CVaR & $2.11$ & $\pm$ & $0.79$ & $5.66$ & $\pm$ & $2.37$ & $26.32$ & $\pm$ & $1.67$ & $2.69$ & $\pm$ & $0.22$ & $0.992$ & $\pm$ & $0.004$\\
\addlinespace[3pt]
Without safety layer & $7.06$ & $\pm$ & $4.18$ & $14.87$ & $\pm$ & $8.88$ & $28.99$ & $\pm$ & $1.55$ & $3.09$ & $\pm$ & $0.25$ & $0.996$ & $\pm$ & $0.003$\\
Without demand margin & $3.31$ & $\pm$ & $1.50$ & $9.51$ & $\pm$ & $5.96$ & $28.47$ & $\pm$ & $1.42$ & $3.06$ & $\pm$ & $0.23$ & $0.997$ & $\pm$ & $0.002$\\
Without CSI margin & $2.64$ & $\pm$ & $1.16$ & $8.38$ & $\pm$ & $4.40$ & $27.16$ & $\pm$ & $1.32$ & $2.90$ & $\pm$ & $0.18$ & $0.996$ & $\pm$ & $0.002$\\
\addlinespace[3pt]
Single policy (no hierarchy) & $5.10$ & $\pm$ & $1.58$ & $22.37$ & $\pm$ & $4.49$ & $3.57$ & $\pm$ & $1.57$ & $0.00$ & $\pm$ & $0.00$ & $0.248$ & $\pm$ & $0.087$\\
\bottomrule
\end{tabular*}
\end{table*}

\subsection{Effect of Carrier Size and Number of Users}\label{subsec:components}\label{subsec:usercount}

\begin{figure}[t]
\centering
\includegraphics[width=\linewidth]{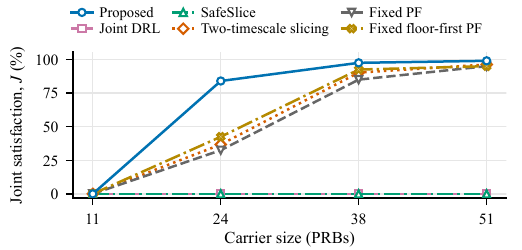}
\caption{$J$ versus carrier size. Joint DRL and SafeSlice remain at 0\%.}
\label{fig:carrier}
\end{figure}

\begin{figure}[t]
\centering
\includegraphics[width=\linewidth]{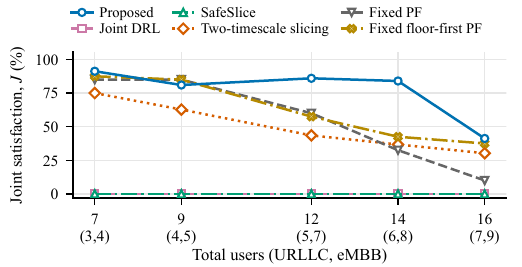}
\caption{$J$ versus user count. Joint DRL and SafeSlice remain at 0\%.}
\label{fig:usercount}
\end{figure}

Fig.~\ref{fig:carrier} compares $J$ across carrier sizes, with users and offered traffic fixed, total power scaling at about 1.667~W per PRB, and the Layer~1 PRB actions and budget limits scaled with the carrier. At 11 PRBs, no method meets both criteria. The advantage of the proposed framework is largest at the nominal 24 PRBs (84\% vs.\ 42.5\% for fixed floor-first PF and 36.75\% for two-timescale slicing) and narrows at 51 PRBs (99\% vs.\ 95\% and 96.25\%), because additional PRBs allow even the fixed schedulers to meet both criteria in most runs.

Fig.~\ref{fig:usercount} compares user counts at 24 PRBs and 40~W, with per-user traffic and service targets unchanged. The
advantage of the proposed framework is largest at the moderate populations $(5,7)$ and $(6,8)$. At light load $(4,5)$, the fixed schedulers achieve higher joint satisfaction (85\% vs.\ 81\%). All failed proposed runs violate the URLLC target despite ample eMBB service. Layer~1 selects its minimum URLLC budget in 87\% of periods, indicating insufficient URLLC reservation at this load. At $(7,9)$, the proposed framework retains an advantage (41.25\% vs.\ 37.5\%), but its worst-user eMBB rate falls below 2~Mbps. Two-timescale slicing provides more aggregate throughput while exceeding the URLLC miss target. Thus, the proposed policy favors URLLC as load increases, while eMBB shortfalls limit joint satisfaction.

\begin{figure}[t]
\centering
\includegraphics[width=\linewidth]{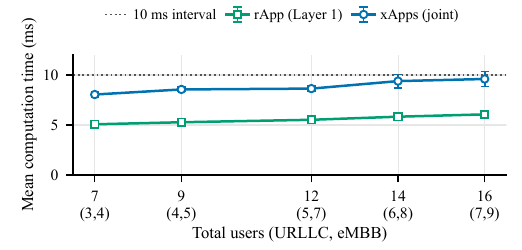}
\caption{Mean rApp and joint xApp decision times. Error bars show one standard deviation across ten training-seed means. The dotted line marks 10~ms.}
\label{fig:runtime_users}
\end{figure}

Furthermore, we measure decision times on an Intel Xeon Gold 6248
(2.50~GHz) using inference with one pinned
CPU thread per model. Fig.~\ref{fig:runtime_users}
shows that the mean joint xApp decision time remains
below 10~ms across all tested user counts.

\subsection{Ablation Study}\label{subsec:ablation}

Table~\ref{tab:ablation} uses ten separately trained models and ten other test environments. The CVaR variants are trained without the removed terms in all four phases, while the safety layer and protection margins are disabled at test time. The single policy selects the PRB owners and the slice power split using the sum of the two assignment rewards, equal power sharing, only the O-DU feasibility checks, and the same network width and training length, and is trained from scratch.

The eMBB CVaR costs raise the worst-user rate by 16\% and the aggregate throughput by 8\% at a higher miss ratio, because they penalize the periods and users with the lowest eMBB rates. To observe more tail runs, we also test the CVaR models on the 40 environments of Table~\ref{tab:nominal}, where, compared with removing all CVaR costs, the complete framework raises $J$ from 74\% to 84.75\% and halves the share of one-second periods with a user below $B_{\min}$ at a similar miss ratio. Removing only the eMBB terms lowers $J$ to 58.25\%, so these gains come from the eMBB terms. The URLLC terms reduce the mean miss ratio by 34\% at a cost of 7\% of the aggregate throughput, but they do not reduce the miss ratio of the worst-served user in the poorest runs (24.54\% with all terms vs.\ 15.14\% without CVaR costs, averaged over the highest 10\% of runs). When the URLLC burst rate is 50\% higher, the eMBB gain persists, whereas the miss ratio with CVaR costs exceeds that without them (11.54\% vs.\ 7.98\%). Hence, the CVaR costs mainly protect eMBB users with low rates and, under URLLC overload, trade URLLC reliability for eMBB rate. The safety layer has the largest test-time effect on URLLC reliability. By transferring 1.33 PRBs per assignment interval on average to URLLC users that fail the service test, it reduces the mean miss ratio by 63\% at a 7.9\% loss of aggregate throughput. Removing the demand margin raises the mean miss ratio from 2.64\% to 3.31\%, because underpredicted bursts then pass the service test with insufficient resources. The demand bound covers every URLLC user in every slot in 90.64\% of calibrated assignment intervals and in 72.29\% at burst onset, above and below $1-\delta=0.9$, respectively. The CSI margin reduces the fraction of scheduled user--slot pairs with overestimated rates from 8.87\% to approximately zero while retaining about 91\% of the rate anticipated from the predicted CSI, and from 19.45\% to 4.02\% when the CSI prediction-error variance is 16 times larger than the calibrated value. Since the model has no decoding errors, an overestimated rate does not cause packet losses. Hence, removing the margin does not raise the miss ratio at either error level and increases aggregate throughput by about 0.47~Mbps. The single policy reduces the aggregate and worst-user eMBB rates to 3.57 and 0.00~Mbps, whereas removing only the safety layer retains 28.99~Mbps, indicating that the hierarchy is essential for eMBB service.

\section{Conclusion}\label{sec:conclusion}

This paper proposed a risk-aware safe hierarchical learning framework for joint eMBB--URLLC resource allocation in O-RAN under traffic-demand and CSI uncertainty. The framework combines two-timescale resource allocation, service-specific CVaR objectives, and an execution-time safety layer using a deterministic per-user URLLC service test with probabilistic guarantees under the stated uncertainty assumptions. At nominal load, the proposed framework was the only evaluated method to meet both the 5\% URLLC deadline-miss target and the 2~Mbps worst-user eMBB rate criterion on average. It achieved a 1.35\% mean miss ratio, representing a 64\% reduction over the most reliable baselines, while delivering 26.16~Mbps aggregate eMBB throughput and the highest worst-user eMBB rate. Ablation results showed that the safety layer reduced URLLC misses by 63\%, while the eMBB CVaR objectives improved joint service satisfaction from 58.25\% to 84.75\%. These findings demonstrate the benefits of combining risk-aware learning with execution-time correction, although performance remains dependent on uncertainty calibration and resource feasibility. Future work will address adaptive uncertainty bounds, multicell coordination, and practical deployment constraints.

\bibliographystyle{IEEEtran}
\bibliography{references}
\end{document}